\documentclass[conference]{IEEEtran}
\IEEEoverridecommandlockouts

\usepackage{cite}
\usepackage{amsmath,amssymb,amsfonts}
\usepackage{algorithmic}
\usepackage{graphicx}
\usepackage{textcomp}
\usepackage{xcolor}
\usepackage{hyperref}
\usepackage{algorithm}
\usepackage{array}
\usepackage[caption=false,font=normalsize,labelfont=sf,textfont=sf]{subfig}
\usepackage{stfloats}
\usepackage{url}
\usepackage{verbatim}
\usepackage{multicol, bm, booktabs, multirow}
\usepackage{relsize}
\allowdisplaybreaks[4]

\def\BibTeX{{\rm B\kern-.05em{\sc i\kern-.025em b}\kern-.08em
    T\kern-.1667em\lower.7ex\hbox{E}\kern-.125emX}}
\begin{document}

\title{Optimal Sensor Placement Using Combinations of Hybrid Measurements for Source Localization
\thanks{© 2025 IEEE.  Personal use of this material is permitted.  Permission from IEEE must be obtained for all other uses, in any current or future media, including reprinting/republishing this material for advertising or promotional purposes, creating new collective works, for resale or redistribution to servers or lists, or reuse of any copyrighted component of this work in other works. DOI: \href{https://doi.org/10.1109/RadarConf2458775.2024.10548509}{10.1109/RadarConf2458775.2024.10548509}.}

}


\author{\IEEEauthorblockN{Kang Tang\textsuperscript{1} , Sheng Xu\textsuperscript{2}, Yuqi Yang\textsuperscript{3}, He Kong\textsuperscript{4}, and Yongsheng Ma\textsuperscript{1} }
\IEEEauthorblockA{\textsuperscript{1} \textit{Shenzhen Key Labor. of Intel. Robo. and Flex. Manuf. Sys., Southern Uni. of Sci. and Tech. (SUSTech), Shenzhen, China} \\
\textsuperscript{2} \textit{Shenzhen Institute of Advanced Technology, Chinese Academy of Sciences, Shenzhen, China} \\
\textsuperscript{3} \textit{Department of Computer Science, University of Nottingham, Nottingham, UK} \\
\textsuperscript{4} \textit{Shenzhen Key Laboratory of Control Theory and Intelligent Systems, SUSTech, Shenzhen, China}\\
Email:12132291@mail.sustech.edu.cn; sheng.xu@siat.ac.cn; psxyy20@nottingham.ac.uk; \{kongh, mays\}@sustech.edu.cn
}}


\maketitle


\begin{abstract}
This paper focuses on static source localization employing different combinations of measurements, including time-difference-of-arrival (TDOA), received-signal-strength (RSS), angle-of-arrival (AOA), and time-of-arrival (TOA) measurements. Since sensor-source geometry significantly impacts localization accuracy, the strategies of optimal sensor placement are proposed systematically using combinations of hybrid measurements. Firstly, the relationship between sensor placement and source estimation accuracy is formulated by a derived Cramér-Rao bound (CRB). Secondly, the A-optimality criterion, i.e., minimizing the trace of the CRB, is selected to calculate the smallest reachable estimation mean-squared-error (MSE) in a unified manner. Thirdly, the optimal sensor placement strategies are developed to achieve the optimal estimation bound. Specifically, the specific constraints of the optimal geometries deduced by specific measurement, i.e., TDOA, AOA, RSS, and TOA, are found and discussed theoretically. Finally, the new findings are verified by simulation studies.
\end{abstract}

\begin{IEEEkeywords}
Optimal sensor placement, hybrid measurements, A-optimality, Cramér-Rao bound, source localization.
\end{IEEEkeywords}

\section{Introduction}
Source localization is important in many applications \cite{c23,c24,Kong2021, Zhou2023IROs,Zhou2023ACC,Zhou2023TRO}, and can be resolved by using diverse sensing modalities and measurements, such as time difference of arrival (TDOA) \cite{c1,Su2021,He2023}, angle of arrival (AOA) \cite{c2}, received signal strength (RSS) \cite{c3,Kong2023}, time of arrival (TOA) \cite{c4}. Typically, the use of multiple sensors can achieve accurate estimation\cite{c6,c7,c8}.

Further, the placement strategies of different sensors have a significant impact on the accuracy improvement of source localization \cite{c5}. To evaluate the localization performance, a quantitative metric is needed. It is well known the Cram\'{e}r-Rao bound (CRB) represents an accuracy limitation for position estimates of an unbiased estimation method, which also builds the relation between the measurement model and the optimal estimation accuracy \cite{c9}. By studying the CRB, the impacts of the sensor geometry on localization accuracy can be acquired.

Many previous studies focused on optimal sensor placement strategies for the scenarios of using individual measurement. The authors in \cite{c10} derived the CRB for TDOA source localization and obtained an optimal deployment strategy. In \cite{c11} and \cite{c12}, the 2-dimensional (2D) and 3-dimensional (3D) optimal placement strategies for the AOA problem were derived. Besides, to solve the localization problems using 2D and 3D RSS and TOA measurements, different solutions were developed in \cite{c13,c14,c15}. In \cite{c16}, the authors have developed a unified framework to quickly calculate the optimal sensor placement for different types of measurements.

The optimal sensor placement problem using hybrid sensing modalities has drawn much attention recently. The work in \cite{c17} offered an initial exploration of the two-group and paired strategies in hybrid TOA-AOA localization, and \cite{c22} provided an optimal sensor deployment strategy for hybrid TOA-AOA localization in two scenarios. The authors in \cite{c18} delved into explicit characterizations of the optimal placement in hybrid TOA, AOA and TDOA localization. The study in \cite{c19} developed optimal sensor placement strategies for the hybrid TOA–RSS–AOA source localization. Although algorithms aimed at specific measurements have been proposed, a unified method to tackle all kinds of hybrid measurements is absent.


This paper focuses on the optimal sensor placement utilizing different combinations of hybrid measurements in terms of TDOA, AOA, RSS and TOA. We theoretically derive the smallest tr(CRB) and propose the optimal sensor placement strategies for all kinds of hybrid measurement combinations in a unified approach, including TDOA-AOA, TDOA-RSS, TDOA-TOA, TDOA-AOA-RSS, TDOA-AOA-TOA, TDOA-RSS-TOA, and TDOA-AOA-RSS-TOA. Furthermore, the constraints for optimal sensor geometries of specific measurements, i.e., TDOA, AOA, RSS, and TOA, in hybrid measurements are revealed and the optimal geometries of all combinations are given in a unified manner. The results are verified through simulations.


\section{Problem Statement}\label{sec:prob}
Consider a stationary source localization problem using $N$ $(N\ge 3)$ sensors on a 2D plane. Each sensor will collect the TDOA, TOA, RSS, and AOA measurements, and thus $N-1$ TDOAs, $N$ TOAs, $N$ RSSs, $N$ AOAs are obtained. Note that when we collect the TDOA measurements, one sensor is regarded as the reference anchor to acquire the $N-1$ TDOA measurements. Let $\bm{s} = {\left( {{s_x},{s_y}} \right)^T}$ denotes the position of the source and ${\bm{p}_i} = {\left( {{x_i},{y_i}} \right)^T}$ is the $i$th sensor. ${d_i} = \left\| {\bm{s} - {\bm{p}_i}} \right\|$ is the distance between the source and the sensor. For simplicity, we assume each type of measurement has a uniform measurement noise variance.

According to~\cite{c18}, the noisy TDOA measurement of sensor $i$ ($i\ge2$) and the reference sensor $1$ takes the form

\begin{equation} \label{tdoa1}
    {\small
    \tilde t_{i1} = \frac{{{d_{i1}}}}{v} + e_{i1}, \quad e_{i1}\sim \mathcal {N}(0,\frac{\sigma^{2}}{v^{2}})\quad}
\end{equation}
where $ d_{i1} = {{d_i}} - {d_1}$ and ${d_i}$ is the distance between the source and sensor $i$. The measurement error $e_{i1}$ is assumed Gaussian white noise (GWN). (\ref{tdoa1}) can be rewritten as
\begin{equation} \label{tdoa}
    \small
    {\tilde d_{i1}} = {d_{i1}} + {ve_{i1}}, \quad ve_{i1}\sim \mathcal {N}(0,\sigma^{2})\quad
\end{equation}
where ${{\tilde d}_{i1}} = v{{\tilde t}_{i1}}$ is the range-difference.

The RSS measurement of the $i$th sensor has \cite{c19}
\begin{equation}
    \small
    \tilde P_{i} = P_{0} - 10\xi {\log _{10}}\left\| {\bm{s} - {\bm{p}_i}} \right\| + {m_i}, \quad m_{i}\sim \mathcal {N}(0,\delta^{2})\quad
\end{equation}
where $P_{0}$ denotes the original signal power at the source and $\xi$ is the path loss parameter depended on the environment. Here the $P_{0}$ and $\xi$ are known and denoted by $A = \frac{-10 \xi}{ln\;10}$. $m_i$ is the measurement error and is assumed GWN.

The noisy AOA measurement is given by \cite{c19}
\begin{equation}
\small
    {\tilde \alpha _i} = {\alpha _i} + {n_i} = {\tan ^{ - 1}}\frac{{{s_y} - {y_i}}}{{{s_x} - {x_i}}} + {n_i}, \quad n_{i}\sim \mathcal {N}(0,\rho^{2})\quad
\end{equation}
where $\alpha_{i}$ is the true azimuth angle between the source and the  $i$th sensor.$n_i$ is the measurement error and is assumed GWN.

The TOA measurement of sensor $i$ is expressed as \cite{c15}
\begin{equation} \label{toa}
\small
    {\tilde g_{i}} = {2g_{i}} + l_{i}, \quad l_{i}\sim \mathcal {N}(0,\gamma^{2})\quad
\end{equation}
where ${g_{i}}$ is the ideal sensor-source distance, and $l_i$ denotes the TOA estimation errors, which is assumed GWN. 

Hence there are $N-1$ TDOAs, $N$ TOAs, $N$ RSSs, and $N$ AOAs.The measurement noise covariance matrix of the $N$ sensors with $4N-1$ measurements are \cite{c20}
{\small
\begin{gather}
    \bm{\Sigma} _{TDOA} = \frac{1}{2}\;\left( {I + {\bf{1}}{{\bf{1}}^T}} \right){\sigma ^2}_{\left( {N - 1} \right) * \left( {N - 1} \right)}, \\
    \bm{\Sigma} _{TOA} = { {{\gamma ^2}I}_{N * N}}, \\
    \bm{\Sigma} _{RSS} = { {{\delta ^2}I}_{N * N}}, \\
    \bm{\Sigma} _{AOA} = { {{\rho ^2}I}_{N * N}}. 
\end{gather}
}

Denote the Jacobian matrix of the $N$ sensors' measurement errors for the source estimation as $J_{TDOA}$, $J_{AOA}$, $J_{RSS}$, $J_{TOA}$, respectively. The $i$th row of the Jacobian matrix is
{\small
\begin{align}
    &\bm{J}_{TDOA}^{i} = \left[ {\frac{{\partial {{\rm{\hat{t}}}_{i1}}}}{{\partial {s_x}}},\frac{{\partial {{\rm{\hat{t}}}_{i1}}}}{{\partial {s_y}}}} \right] = \left[ {\cos {\alpha _i} - \cos {\alpha _1},\sin {\alpha _i} - \sin {\alpha _1}} \right],\\
    &\bm{J}_{AOA}^{i} = \left[ {\frac{{\partial {{\rm{\hat{\alpha}}}_{i}}}}{{\partial {s_x}}},\frac{{\partial {{\rm{\hat{\alpha}}}_{i}}}}{{\partial {s_y}}}} \right] = \left[ {\frac{-\sin {\alpha _i}}{d_{i}},\frac{\cos {\alpha _i}}{d_{i}}} \right],\\
    &\bm{J}_{RSS}^{i} = \left[ {\frac{{\partial {{\rm{\hat{P}}}_{i}}}}{{\partial {s_x}}},\frac{{\partial {{\rm{\hat{P}}}_{i}}}}{{\partial {s_y}}}} \right] = \left [ {\frac{\cos {\alpha _i}}{d_{i}},\frac{\sin {\alpha _i}}{d_{i}}} \right],\\
    &\bm{J}_{TOA}^{i} = \left[ {\frac{{\partial {{\rm{\hat{g}}}_{i}}}}{{\partial {s_x}}},\frac{{\partial {{\rm{\hat{g}}}_{i}}}}{{\partial {s_y}}}} \right] = \left[ {\cos {\alpha _i},\sin {\alpha _i}} \right].
\end{align}
}

In terms of the estimation of the source position $\bm{s}$, the CRB gives the lower bound for the covariance matrix of an unbiased estimated state $\hat{\bm{s}}$ \cite{c9}
\begin{equation}
\small
    \mathbb{E}\left[ {\left( {{\bm{s}} - \hat{\bm{s}}} \right){{\left( {{\bm{s}} - \hat{\bm{s}}} \right)}^T}} \right] \succeq \bm{F}^{-1}
\end{equation}
where $\bm{F}$ is the Fisher information matrix (FIM) and CRB is the inverse of the FIM. It is proved that minimizing the tr(CRB) is equivalent to minimizing the estimation MSE \cite{c21}. In this paper, we will use this lower bound as a criterion for the sensor-source geometry optimization.

\section{Optimal Sensor Placement Strategies}
\label{sec:opt_1}

In our system, every sensor comprises different mechanisms to collect the TDOA, TOA, RSS, and AOA measurements. Therefore, it is reasonable to assume the noise in the different measurements to be independent of each other. Considering different combinations of hybrid measurements, we will derive the tr(CRB)s using hybrid TDOA-AOA-RSS-TOA measurements firstly, and then other hybrid measurements including TDOA-RSS, TDOA-AOA, TDOA-TOA, TDOA-AOA-TOA, TDOA-RSS-TOA, and TDOA-AOA-RSS, will be analyzed. 


\subsection{Hybrid TDOA-AOA-RSS-TOA Measurements}
\label{ssec:hyb_1}

Firstly, we consider the hybrid TDOA-AOA-RSS-TOA measurements in which each sensor can obtain a TDOA, an AOA, an RSS, and a TOA measurements simultaneously. Using $\bm{\Sigma}$ to denote the noise covariance matrix and we have
\begin{equation}
\small
    \mathbf{\Sigma}  = \text{diag}\left( {{\mathbf{\Sigma} _{TDOA}},{\mathbf{\Sigma} _{AOA}},{\mathbf{\Sigma} _{RSS}},{\mathbf{\Sigma} _{TOA}}} \right).
\end{equation}
the FIM and its CRB can be written as \cite{c22}
\begin{equation}\label{fim7}
\small
\begin{split}
\bm{F} &= \bm{J}{\bm{\Sigma} ^{ - 1}}{\bm{J}^T}  \\
&= {\bm{J}_{TDOA}}\bm{\Sigma} _{TDOA}^{ - 1}\bm{J}_{TDOA}^T + {\bm{J}_{TOA}}\bm{\Sigma} _{TOA}^{ - 1}\bm{J}_{TOA}^T \\
&+ {\bm{J}_{RSS}}\bm{\Sigma} _{RSS}^{ - 1}\bm{J}_{RSS}^T + {\bm{J}_{AOA}}\bm{\Sigma} _{AOA}^{ - 1}\bm{J}_{AOA}^T
\end{split}
\end{equation}
where CRB = ${\bm{F} ^{ - 1}}$. Since it is a two-dimensional problem, $F$ is a symmetrical and semi-positive matrix that has a size of $2\times 2$ and it can be represented as follows \cite{c21}.
\begin{equation}
\small
    {\bm{F} = \left[ {\begin{array}{*{20}{c}}
{{\varphi _{11}}}&{{\varphi _{12}}}\\
{{\varphi _{21}}}&{{\varphi _{11}}}
\end{array}} \right]}
\end{equation}
where $\varphi_{12} = \varphi_{21}$. The details of FIM are derived in (\ref{long form}).

\begin{figure*}[ht] 
\small
\centering 
\vspace*{8pt} 
\begin{gather} 
    \varphi_{11} = {\sum\limits_{i = 1}^N {\left[ {\left( {\frac{1}{{{\sigma ^2}}} + \frac{4}{{{\gamma ^2}}} + \frac{{{A^2}}}{{{\delta ^2}d_i^2}}} \right){{\cos }^2}{\alpha _i} + \frac{1}{{{\rho ^2}d_i^2}}{{{\sin }^2}{\alpha _i}}} \right]}  \notag - \frac{1}{{N{\sigma ^2}}}{{\left( {\sum\limits_{i = 1}^N {\cos {\alpha _i}} } \right)}^2}}, \\ \label{long form}
    \varphi_{12} = {\sum\limits_{i = 1}^N {\left[ {\frac{1}{{{\sigma ^2}}} + \frac{4}{{{\gamma ^2}}} + \left( {\frac{{{A^2}}}{{{\delta ^2}}} - \frac{1}{{{\rho ^2}}}} \right)\frac{1}{{d_i^2}}} \right]\cos {\alpha _i}\sin {\alpha _i}}  - \frac{1}{{N{\sigma ^2}}}\sum\limits_{i = 1}^N {\cos {\alpha _i}} \sum\limits_{i = 1}^N {\sin {\alpha _i}} }, \\
    \varphi_{22} = {\sum\limits_{i = 1}^N {\left[ {\left( {\frac{1}{{{\sigma ^2}}} + \frac{4}{{{\gamma ^2}}} + \frac{{{A^2}}}{{{\delta ^2}d_i^2}}} \right){{\sin }^2}{\alpha _i} + \frac{1}{{{\rho ^2}d_i^2}}{{{\cos }^2}{\alpha _i}}} \right]}  - \frac{1}{{N{\sigma ^2}}}{{\left( {\sum\limits_{i = 1}^N {\sin {\alpha _i}} } \right)}^2}}. \nonumber 
\end{gather}
\hrulefill 
\begin{equation}
\label{crb_gen}
    \small
\begin{split}
O_{1} = \frac{4}{{\frac{N}{{{\sigma ^2}}} + \frac{1}{{{\rho ^2}}}\sum\limits_{i = 1}^N {\frac{1}{{d_i^2}}} }}, \;\;\;\;
O_{2} = \frac{4}{{\frac{N}{{{\sigma ^2}}} + \frac{{{A^2}}}{{{\delta ^2}}}\sum\limits_{i = 1}^N {\frac{1}{{d_i^2}}} }},  \;\;\;\;\;
O_{3} = \frac{4}{(\frac{1}{{{\sigma ^2}}} + \frac{4}{{{\gamma ^2}}})N} \;\;\;\;
O_{4} = \frac{4}{{\frac{1}{{{\sigma ^2}}}N + (\frac{1}{{{\rho ^2}}} + \frac{{{A^2}}}{{{\delta ^2}}})\sum\limits_{i = 1}^N {\frac{1}{{d_i^2}}} }},\\
O_{5} = \frac{4}{{(\frac{1}{{{\sigma ^2}}} + \frac{4}{{{\gamma ^2}}})N + \frac{{{A^2}}}{{{\delta ^2}}}\sum\limits_{i = 1}^N {\frac{1}{{d_i^2}}} }},\;\;\;\;
O_{6} = \frac{4}{{(\frac{1}{{{\sigma ^2}}} + \frac{4}{{{\gamma ^2}}})N + \frac{1}{{{\rho ^2}}}\sum\limits_{i = 1}^N {\frac{1}{{d_i^2}}} }},\;\;\;\;
O_{7} = \frac{4}{{(\frac{1}{{{\sigma ^2}}} + \frac{4}{{{\gamma ^2}}})N + (\frac{1}{{{\rho ^2}}} + \frac{{{A^2}}}{{{\delta ^2}}})\sum\limits_{i = 1}^N {\frac{1}{{d_i^2}}} }}.
\end{split}
\end{equation}
\hrulefill 
\end{figure*}

According to \cite{c15}, there is an inequality that the tr(CRB) cannot be smaller than the sum of the FIMs reciprocal diagonal elements. Denote that $B = \sum_{i = 1}^N { {\cos {\alpha _i}}}$, $C = \sum_{i = 1}^N {\sin {\alpha _i}}$, $D = \sum_{i = 1}^N {{{\left( {\cos {\alpha _i}} \right)}^2}}$, $E = \sum_{i = 1}^N {{{\left( {\sin {\alpha _i}} \right)}^2}}$, $K = \sum_{i = 1}^N {\frac{{{{\left( {cos{\alpha _i}} \right)}^2}}}{{d_i^2}}}$, $G = \sum_{i = 1}^N {\frac{{{{\left( {sin{\alpha _i}} \right)}^2}}}{{d_i^2}}}$, $H = {\frac{1}{{{\sigma ^2}}} + \frac{4}{{{\gamma ^2}}}}$, $L = {\frac{1}{{{\rho ^2}}} + \frac{{{A^2}}}{{{\delta ^2}}}}$. Therefore, we can obtain
\begin{equation}
\small
\begin{split}
\text{tr}\left( {{{\bm{F }}^{-1}}} \right)& \ge \frac{1}{{{\varphi _{11}}}} + \frac{1}{{{\varphi _{22}}}} \\
&= \frac{1}{{\left( {\frac{1}{{{\sigma ^2}}} + \frac{4}{{{\gamma ^2}}}} \right)D - \frac{1}{{N{\sigma ^2}}}{B^2} + \frac{1}{{{\rho ^2}}}G + \frac{{{A^2}}}{{{\delta ^2}}}F}} \\
&+ \frac{1}{{\left( {\frac{1}{{{\sigma ^2}}} + \frac{4}{{{\gamma ^2}}}} \right)E - \frac{1}{{N{\sigma ^2}}}{C^2} + \frac{1}{{{\rho ^2}}}K + \frac{{{A^2}}}{{{\delta ^2}}}G}} \\
\end{split}
\end{equation}
subject to $F$ or $F^{-1}$ is diagonal. The equality constraints will be discussed later. Moreover, we have
\begin{equation}
\small
\begin{split}
\text{tr}\left( {{{\bm{F }}^{ -1}}} \right) &\ge \frac{1}{{{\varphi _{11}}}} + \frac{1}{{{\varphi _{22}}}} \ge \frac{4}{{{\varphi _{11}} + {\varphi _{22}}}} \\
&= \frac{4}{HN + L\sum\limits_{i = 1}^N {\frac{1}{{d_i^2}}}  - \frac{1}{{N{\sigma ^2}}}\left( {{B^2} + {C^2}} \right)} \\
&\ge \frac{4}{HN + {L\sum\limits_{i = 1}^N {\frac{1}{{d_i^2}}} }}\\
\end{split}
\end{equation}
subject to
{
\small
\begin{gather}\label{cons_11}
    {\varphi _{12}} = {\varphi _{21}} = 0 \\
    {\varphi _{11}} = {\varphi _{22}}   \label{cons_12} \\ \label{cons_13}
    {B^2} + C^2 = 0
\end{gather}}%
where we apply the inequality transformation, i.e., $\frac{1}{a} + \frac{1}{b} \ge \frac{4}{{a + b}}$ when $a \ge 0$, $b \ge 0$ and the equality hold if $a = b$. Therefore, the lower bound of the tr(CRB) is obtained analytically and denote the optimal value as $O_{7}$, and we have
\begin{equation} \label{CRB_TDOA-AOA}
\small
    \text{tr}\left( {\text{CRB}} \right) \ge \frac{4}{{(\frac{1}{{{\sigma ^2}}} + \frac{4}{{{\gamma ^2}}})N + (\frac{1}{{{\rho ^2}}} + \frac{{{A^2}}}{{{\delta ^2}}})\sum\limits_{i = 1}^N {\frac{1}{{d_i^2}}} }} \triangleq O_{7}.
\end{equation}

Next, we investigate the geometrical constraints corresponding to (\ref{cons_11})-(\ref{cons_13}). Since each sensor collects various types of measurements, the corresponding measurement variances exhibit different units. Therefore, we assume that different types of variances, e.g., $\frac{1}{{{\sigma ^2}}}\sum\limits_{i = 1}^N {{{\left( {\cos {\alpha _i}} \right)}^2}}$ and $\frac{4}{{{\gamma ^2}}}\sum\limits_{i = 1}^{{N}} {{{\left( {\cos {\alpha _i}} \right)}^2}}$, are not additive \cite{c19}, when deriving the optimal sensor placement. In order to satisfy (\ref{cons_13}), we have
\begin{equation}
\small
    \sum\limits_{i = 1}^N {\sin {\alpha _i}} = 0\;\;\text{and}\;\;\sum\limits_{i = 1}^N {\cos {\alpha _i}} = 0.  \label{op-1}
\end{equation}
Then substituting (\ref{op-1}) into (\ref{cons_11}) and (\ref{cons_12}), the constraints are transformed to
{\small
\begin{gather}
\frac{1}{{{\sigma ^2}}}\sum\limits_{i = 1}^N {{{\left( {\cos {\alpha _i}} \right)}^2}}  = \frac{1}{{{\sigma ^2}}}\sum\limits_{i = 1}^N {{{\left( {\sin {\alpha _i}} \right)}^2}}  \Rightarrow \sum\limits_{i = 1}^N {\cos 2{\alpha _i}}  = 0\\
\frac{1}{{{\rho ^2}}}\sum\limits_{i = 1}^N {\frac{{{{\left( {\cos {\alpha _i}} \right)}^2}}}{{d_i^2}}}  = \frac{1}{{{\rho ^2}}}\sum\limits_{i = 1}^N {\frac{{{{\left( {\sin {\alpha _i}} \right)}^2}}}{{d_i^2}}}  \Rightarrow \sum\limits_{i = 1}^N {\frac{{\cos 2{\alpha _i}}}{{d_i^2}}}  = 0 \\
\frac{1}{{{\sigma ^2}}}\sum\limits_{i = 1}^N {\cos {\alpha _i}\sin {\alpha _i}}  = 0 \Rightarrow \sum\limits_{i = 1}^N {\sin 2{\alpha _i}}  = 0\\
\frac{1}{{{\rho ^2}}}\sum\limits_{i = 1}^N {\frac{{\sin {\alpha _i}\cos {\alpha _i}}}{{d_i^2}}}  = 0 \Rightarrow \sum\limits_{i = 1}^N {\frac{{\sin 2{\alpha _i}}}{{d_i^2}}}  = 0.
\end{gather}
}

Finally, the optimal placement solution is expressed as
{\small
\begin{gather}
\sum\limits_{i = 1}^N {\sin {\alpha _i}} = 0,\;\;\sum\limits_{i = 1}^N {\cos {\alpha _i}} = 0,  \label{op-1-1} \\
\sum\limits_{i = 1}^N {\sin 2{\alpha _i}} = 0,\;\;\sum\limits_{i = 1}^N {\cos 2{\alpha _i}} = 0,  \label{op-1-2}\\
\sum\limits_{i = 1}^N {\frac{{\sin 2{\alpha _i}}}{{d_i^2}}}  = 0,\;\;\sum\limits_{i = 1}^N {\frac{{\cos 2{\alpha _i}}}{{d_i^2}}}  = 0. \label{op-1-3}
\end{gather}
}

Since it's a complex non-convex problem to find the solution to (\ref{op-1-1}) - (\ref{op-1-3}), numerical results will be investigated to present final optimal geometrical strategies later.

\subsection{Other Cases}
\label{sssec:hyb_2}
In addition to the hybrid TDOA-AOA-RSS-TOA problem, there are many other combinations of the hybrid measurements. In this section, we calculate the tr(CRB)s of TDOA-based hybrid measurements problems in the same manner, including TDOA-RSS, TDOA-AOA, TDOA-TOA, TDOA-AOA-TOA, TDOA-RSS-TOA, and TDOA-AOA-RSS measurements. After some derivations which are similar to section 3.1, the smallest tr(CRB)s of these combinations mentioned above are derived in (\ref{crb_gen}), and they are denoted as $O_{1},O_{2},O_{3},O_{4},O_{5},O_{6}$, respectively.

For example, the corresponding optimal placement solutions of hybrid TDOA-TOA source localization are given by 
{\small
\begin{gather}
\sum\limits_{i = 1}^N {\sin {\alpha _i}} = 0,\;\;\sum\limits_{i = 1}^N {\cos {\alpha _i}} = 0,  \label{op-1-4} \\
\sum\limits_{i = 1}^N {\sin 2{\alpha _i}} = 0,\;\;\sum\limits_{i = 1}^N {\cos 2{\alpha _i}} = 0,  \label{op-1-5}
\end{gather}
}%
where the detailed derivations are omitted. Comparing with the solution in (\ref{op-1-1})-(\ref{op-1-3}), there is a little difference between these two scenarios. More specifically, the optimal geometry for the hybrid TDOA-TOA problem has four constraints, but the constraint number is six in the hybrid TDOA-AOA-RSS-TOA problem.

Moving forward, all the other combinations of hybrid measurements source localization, i.e. TDOA - RSS, TDOA - AOA - RSS, TDOA - AOA - TOA, TDOA - RSS - TOA, and TDOA - AOA - RSS, have the completely same optimal solutions as the results of the hybrid TDOA-AOA-RSS-TOA problem in (\ref{op-1-1})-(\ref{op-1-3}).

\subsection{Observations}
\label{sssec:obv_1}

Based on the optimal solutions of the measurements combinations (\ref{op-1-1})-(\ref{op-1-5}), we aim to investigate the relationship between the different solutions of hybrid source localization and the fourth measurements, i.e., TDOA, AOA, RSS, and TOA, which means that we try to find the specific constraints generated by each kind of measurement in a hybrid scenario.

Notice that the optimal geometry of hybrid TDOA-TOA measurements in (\ref{op-1-4})-(\ref{op-1-5}) has two fewer constraints than that of hybrid TDOA-AOA-TOA measurements in (\ref{op-1-1})-(\ref{op-1-3}). From this observation, we can infer that the constraints (\ref{op-1-3}) are generated by AOA measurements specifically.

Similarly, we can find that the constraints (\ref{op-1-3}) concurrently belong to the RSS measurements because the optimal solutions of hybrid TDOA-AOA-TOA and hybrid TDOA-RSS-TOA are absolutely the same in (\ref{op-1-1})-(\ref{op-1-3}).
\\
\emph{Observation 1.} Based on the interpretation above, it can be deduced that the AOA measurements and the RSS measurements generate the same constraints (\ref{op-1-3}) in the optimal placement solutions of hybrid measurements.
\\
\emph{Observation 2.} From another perspective, comparing the constraints of hybrid TDOA-TOA case in (\ref{op-1-4})-(\ref{op-1-5}) and the TDOA-AOA case in (\ref{op-1-1})-(\ref{op-1-3}), the constraints (\ref{op-1-1})-(\ref{op-1-2}) are generated by the TDOA measurements and the constraints responding to TOA are part of the TDOA. It seems that the constraints of TOA are concluded by TDOA's constraints when they appear simultaneously.
\\
\emph{Observation 3.} In addition, we find that the the constraints related to TDOA and TOA are determined by the azimuth angles between the source and the sensors while the constraints caused by AOA and RSS are determined by both the azimuth angles and the source-sensor distances. 

All in all, the optimal geometry solutions can be rewritten in a unified form as
{\small
\begin{align}
&\left. {\begin{array}{*{20}{c}}
\sum\limits_{i = 1}^N {\sin {\alpha _i}} = 0,\;\;\;\;\sum\limits_{i = 1}^N {\cos {\alpha _i}} = 0, \\
\sum\limits_{i = 1}^N {\sin 2{\alpha _i}} = 0,\;\;\sum\limits_{i = 1}^N {\cos 2{\alpha _i}} = 0, \label{op-1-6}
\end{array}} \right\}TDOA, \\
&\left. {\begin{array}{*{20}{c}}
\sum\limits_{i = 1}^N {\sin 2{\alpha _i}} = 0,\;\;\sum\limits_{i = 1}^N {\cos 2{\alpha _i}} = 0, \label{op-1-8}
\end{array}} \right\}TOA, \\
&\left. {\begin{array}{*{20}{c}}
\sum\limits_{i = 1}^N {\frac{{\sin 2{\alpha _i}}}{{d_i^2}}}  = 0,\;\;\sum\limits_{i = 1}^N {\frac{{\cos 2{\alpha _i}}}{{d_i^2}}}  = 0, \label{op-1-7}
\end{array}} \right\}AOA/RSS.
\end{align}
}
\\
\emph{Observation 4}. The specific measurement in the hybrid source localization plays a specific role in the optimal sensor placement. If we need to acquire the optimal sensor-source geometry using a part of the four types of measurements, we can make a combination using the solution in (\ref{op-1-6})-(\ref{op-1-7}) easily.

\section{Simulation Results}
\label{sec:sim}
Since the geometries are related to both the azimuth angles and the sensor-source distances, we verify the solutions by different examples using both uniform and nonuniform $d_{i}$.


\subsection{Results Under Uniform $d_i$}

In some real applications, such as underwater searching and radar systems, the sensors are not close to the source. In many scenarios, multiple mobile platforms such as Unmanned Aerial Vehicle (UAV) or Unmanned Ground Vehicle (UGV) carrying the sensors are employed to localize the source. Consider that some algorithms will be used to estimate an initial source position in advance. Hence, the movable vehicles can plan the path and move to some specific distance $d_{i}$. Due to the motion constraints of the sensor platforms, such as space or safety concerns, the sensors could not get too close to the source. 

Since page limitation, we only provide the simulations in a representative scenario using the hybrid TDOA-AOA-RSS-TOA measurements. In addition, all the sensors move to a uniform $d_{i} = d_{0}$, then the geometries constraints in (\ref{op-1-1})-(\ref{op-1-3}) are simplified to (\ref{op-1-4})-(\ref{op-1-5}).

\emph{Results}. One of the unified solutions of (\ref{op-1-4}) - (\ref{op-1-5}) is uniform angular array (UAA) under (i) uniform $d_{i}=d_{0}$, i.e., the azimuth angle $\alpha_{i}$ have the equal angular distribution. (ii) large $d_{i}$, e.g., $d_{i}\ge 100$ m.

\begin{figure}[ht]
\begin{minipage}[b]{0.49\linewidth}
  \centering
  \centerline{\includegraphics[width=4.5cm]{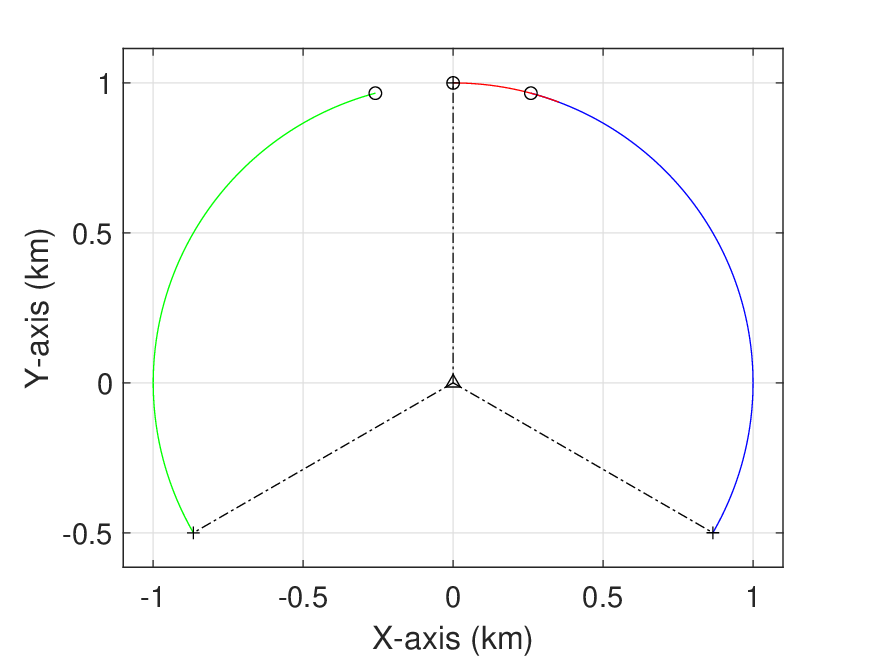}}
\end{minipage}
\hfill
\begin{minipage}[b]{0.49\linewidth}
  \centering
  \centerline{\includegraphics[width=4.5cm]{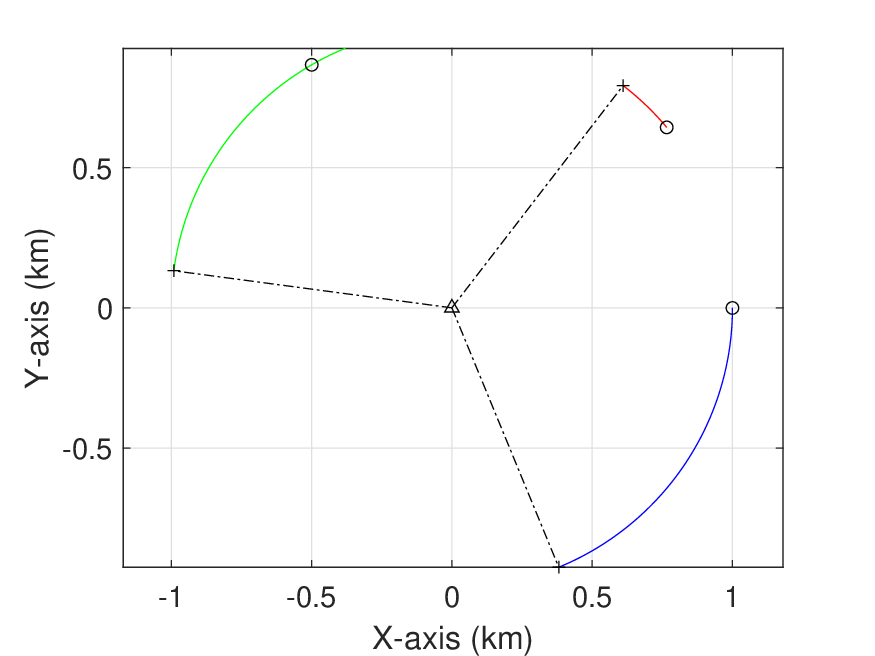}}
\end{minipage}
\begin{minipage}[b]{0.49\linewidth}
  \centering
  \centerline{\includegraphics[width=4.5cm]{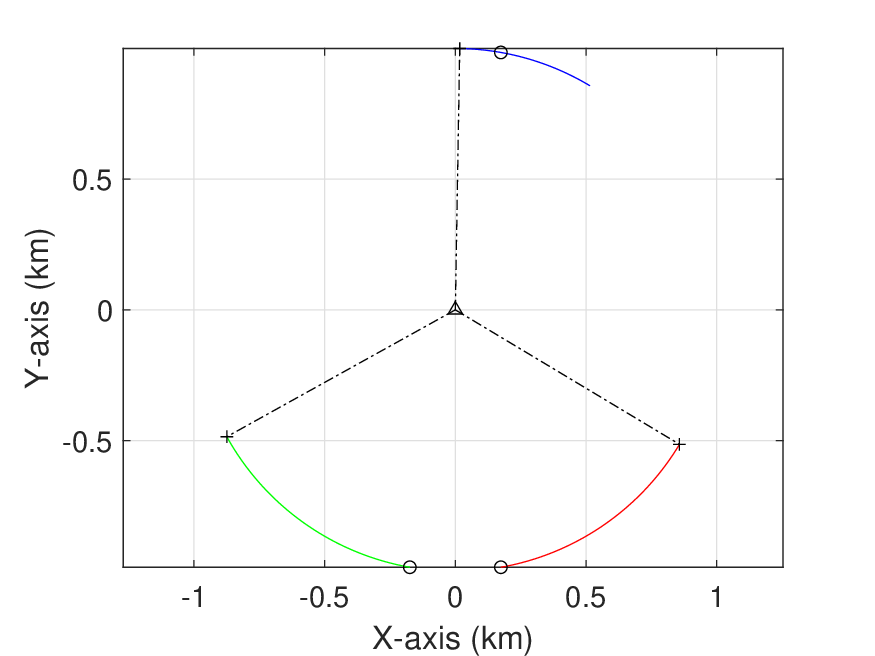}}
\end{minipage}
\hfill
\begin{minipage}[b]{0.49\linewidth}
  \centering
  \centerline{\includegraphics[width=4.5cm]{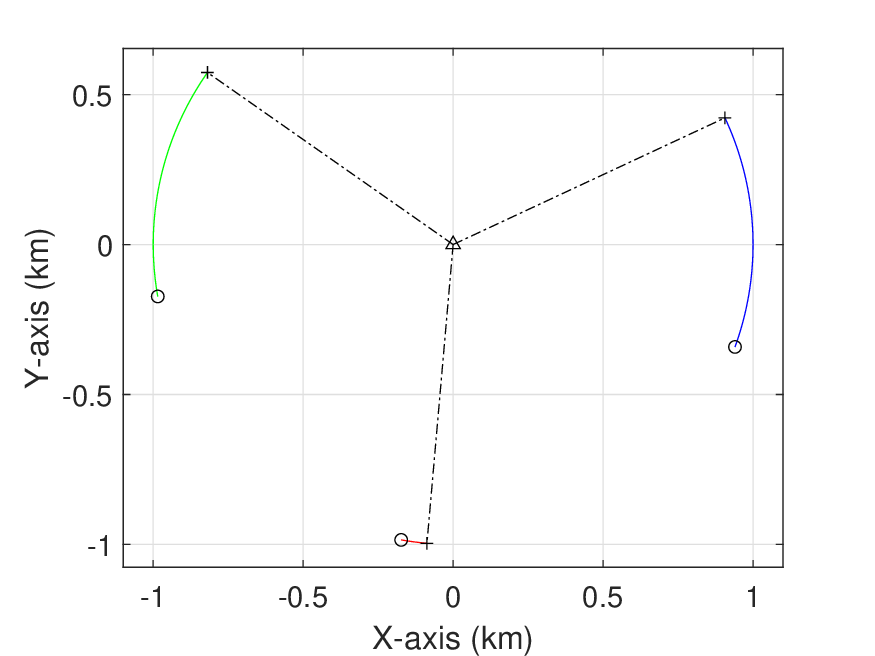}}
\end{minipage}
\caption{Optimal sensor trajectories of three sensors with uniform $d_{i} = 1000$m. Each black $\circ$ denotes the start position of a sensor, and a black $+$ denotes the final position of a sensor. The true source is denoted by the black $\triangle$. The dashed lines indicate the final optimal sensor-source geometries.}
\label{fig_uni_3}
\end{figure}

\emph{Case 1 (uniform)}: Assume that $N=3$, $P_{0} = 1000$W, $\xi = 1$W/m, $\sigma = 0.5$m, $\rho = 1^{\circ}$, $\delta = 1$W,  $\gamma = 1.5$m, and $d_{i} = d_{0} = 1000$m for any $i$. The gradient descent algorithm that aims to search the minimal tr(CRB) is employed to investigate the optimal geometrical strategies \cite{c19}. Generally, the gradient-based algorithm uses 10000 steps. Three mobile sensors move from random positions. As shown in \textbf{Fig}. \ref{fig_uni_3}, in order to inspect the robustness of the strategies, the simulations are divided into four examples with different initial positions of the sensors, i.e., {$\{\alpha_{1} = 75^{\circ}, \;\alpha_{2} = 90^{\circ}, \;\alpha_{3} = 105^{\circ}\}$}, {$\{\alpha_{1} = 0^{\circ}, \;\alpha_{2} = 40^{\circ}, \;\alpha_{3} = 120^{\circ}\}$}, {$\{\alpha_{1} = 80^{\circ}, \;\alpha_{2} = -80^{\circ}, \;\alpha_{3} = -100^{\circ}$\}}, {$\{\alpha_{1} = -20^{\circ}, \;\alpha_{2} = -100^{\circ}, \;\alpha_{3} = 190^{\circ}$\}}. respectively. As a result, all the final sensor-source geometries of the four examples satisfy the constraints in (\ref{op-1-4})-(\ref{op-1-5}), which indicates that they all achieve the theoretical optimal geometries. The final optimal geometries are: {$\{\alpha_{1} = 150.01^{\circ}, \;\alpha_{2} = -89.99^{\circ}, \;\alpha_{3} = 30.00^{\circ}\}$}, {$\{\alpha_{1} = 112.39^{\circ}, \;\alpha_{2} = -127.61^{\circ}, \;\alpha_{3} = -7.60^{\circ}\}$}, {$\{\alpha_{1} = -90.97^{\circ}, \;\alpha_{2} = 149.03^{\circ}, \;\alpha_{3} = 29.04^{\circ}$\}}, {$\{\alpha_{1} = -155.01^{\circ}, \;\alpha_{2} = 84.99^{\circ}, \;\alpha_{3} = -35.01^{\circ}$\}}, respectively. The simulation results indicate that the optimal placement strategies satisfy the UAA condition and reach the corresponding smallest tr(CRB) = 0.2306$\rm m^{2}$. 

\begin{figure}[ht]
\begin{minipage}[b]{0.45\linewidth}
  \centering
  \centerline{\includegraphics[width=4.5cm]{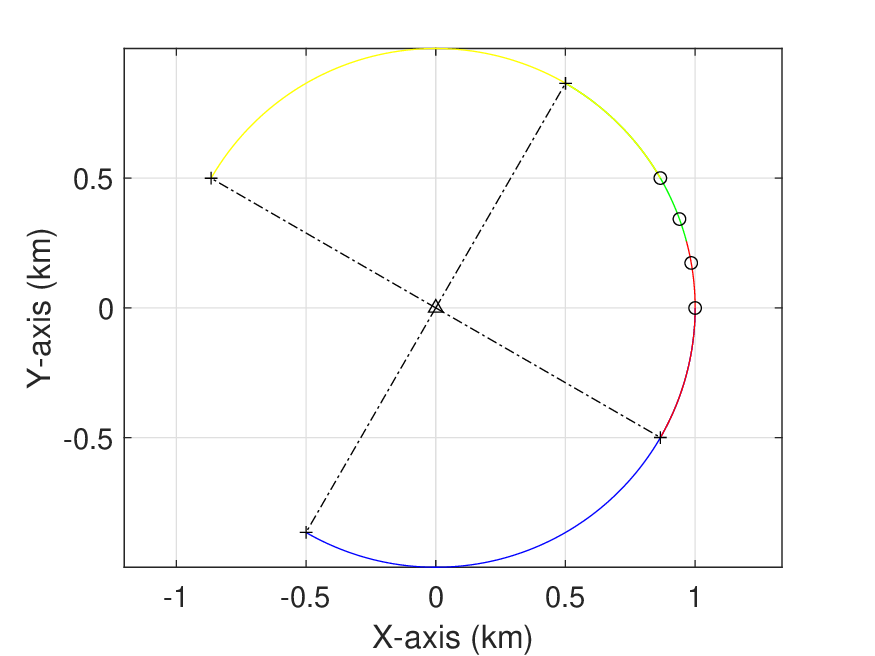}}
\end{minipage}
\hfill
\begin{minipage}[b]{0.45\linewidth}
  \centering
  \centerline{\includegraphics[width=4.5cm]{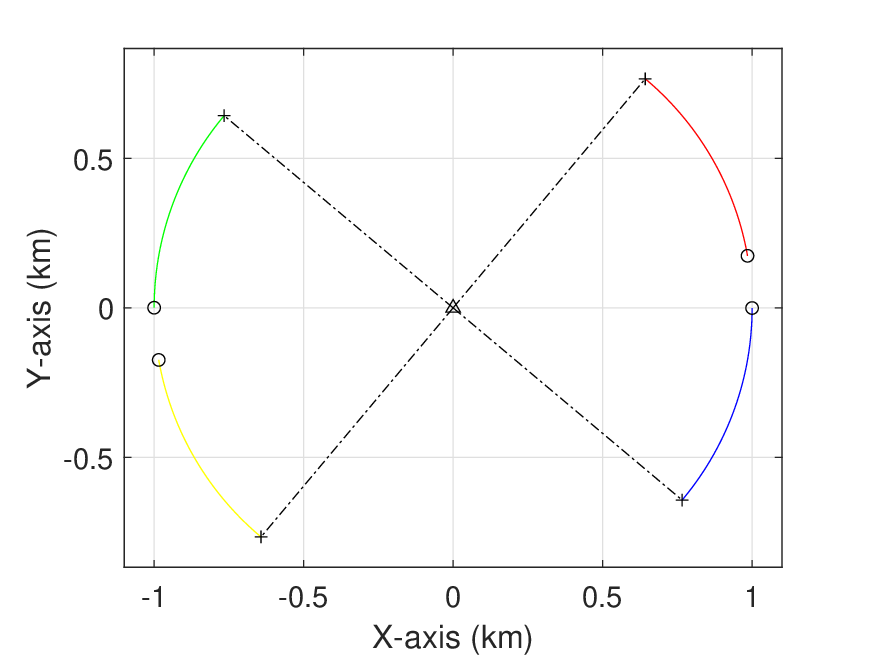}}
\end{minipage}
\caption{Optimal sensor trajectories with different 4 start points for uniform $d_{i} = 1000\;m$}
\label{fig_uni_4}
\end{figure}

\emph{Case 2 (uniform)}: The parameters are assumed that $N = 4$, $P_{0} = 1000\;W$, $\xi = 1\;W/m$, $\sigma = 1\;m$, $\rho = 2^{\circ}$,$\delta = 2\; W$,  $\gamma = 2\; m$, and $d_{i} = d_{0} = 1000\;m$ for any $i$. The initial positions of the 4 sensors are {$\{\alpha_{1} = 0^{\circ}, \;\alpha_{2} = 10^{\circ}, \;\alpha_{3} = 20^{\circ}, \;\alpha_{4} = 30^{\circ}\}$}, {$\{\alpha_{1} = 0^{\circ}, \;\alpha_{2} = 10^{\circ}, \;\alpha_{3} = 180^{\circ}, \;\alpha_{4} = 190^{\circ}\}$}, respectively. The simulation result are demonstrated in \textbf{Fig}. 2 after applying the same approach in \emph{Case 1}. The optimal tr(CRB) is calculated as $0.4998\;m^{2}$ and the final positions of the sensors are UAA. It is obvious that the in each sensor group the optimal geometries fit the analytical solutions well and the strategies are also UAA.

\begin{figure}[ht]
\begin{minipage}[b]{0.45\linewidth}
  \centering
  \centerline{\includegraphics[width=4.5cm]{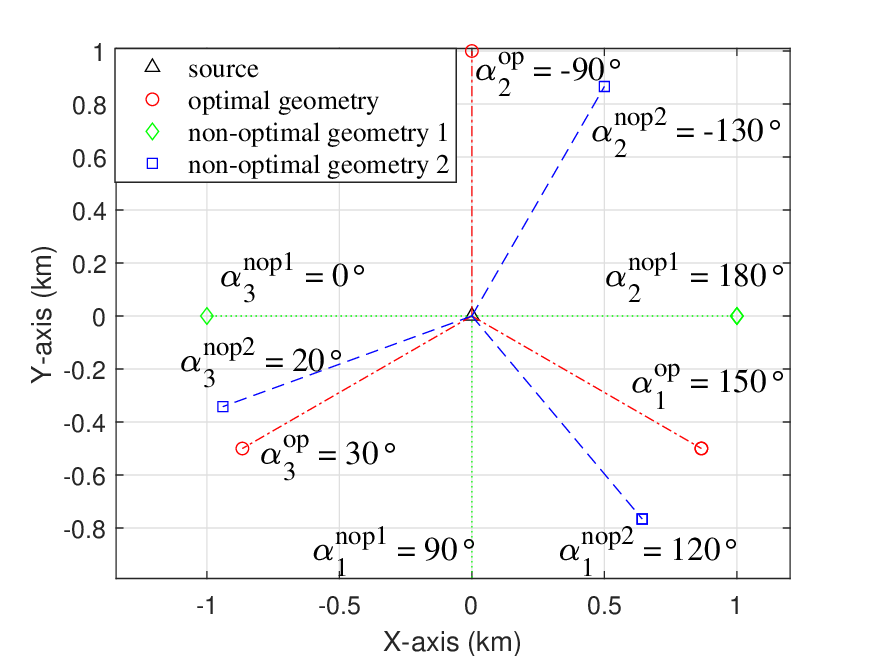}}
  \centerline{(a)}
\end{minipage}
\hfill
\begin{minipage}[b]{0.45\linewidth}
  \centering
  \centerline{\includegraphics[width=4.5cm]{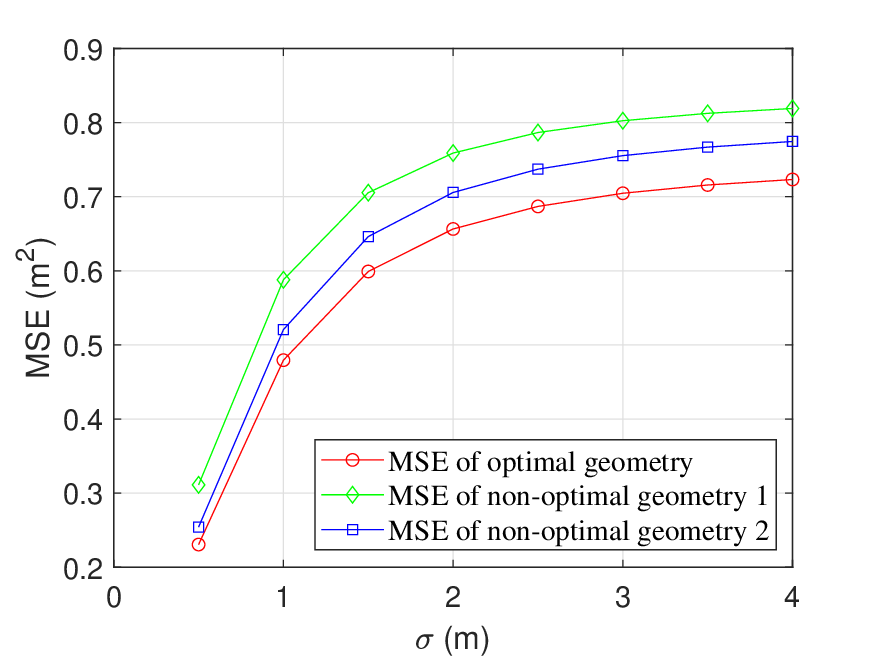}}
  \centerline{(b)}
\end{minipage}
\caption{Comparison of optimal and non-optimal strategies, (a) strategies and (b) MSEs as a function of different noise error $\sigma$.}
\label{fig_comp}
\end{figure}

\emph{Case 3 (uniform)}: In order to inspect the estimation accuracy, we compare one optimal geometry with two non-optimal geometries, and the parameters are the same as those of \textbf{Fig}. 1(a). Different noise errors of standards are utilized to show the mean-squared-errors (MSEs), i.e., we use MSEs of the different geometries as a function of different $\sigma$ values (for simplification, we use different $\sigma$). As shown in \textbf{Fig}. 3, the optimal strategy can obtain a better accuracy than the non-optimal ones.

\subsection{Results under Nonuniform $d_i$}

The mobile platforms are always restricted to the workspace and sometimes we need to place the sensors in nonuniform circumstances. When $d_{i}$ is large, the constraints $\sum_{i = 1}^N {\frac{{\sin 2{\alpha _i}}}{{d_i^2}}}  = 0$  and $\sum_{i = 1}^N {\frac{{\cos 2{\alpha _i}}}{{d_i^2}}}  = 0$ are approximate to the constraints $\sum_{i = 1}^N {\sin 2{\alpha _i}} = 0$ and $\sum_{i = 1}^N {\cos 2{\alpha _i}} = 0$ but if $d_{i}$ is comparable to the $\cos 2{\alpha _i}$, the constraints in (\ref{op-1-1})-(\ref{op-1-3}) should be satisfied strictly. Hence, when $d_{i}$ is very large, the optimal geometries should have a similar form to the UAA condition, otherwise, the geometries will depend on $d_{i}$.
\begin{figure}[htb]
\begin{minipage}[b]{0.45\linewidth}
  \centering
  \centerline{\includegraphics[width=4.5cm]{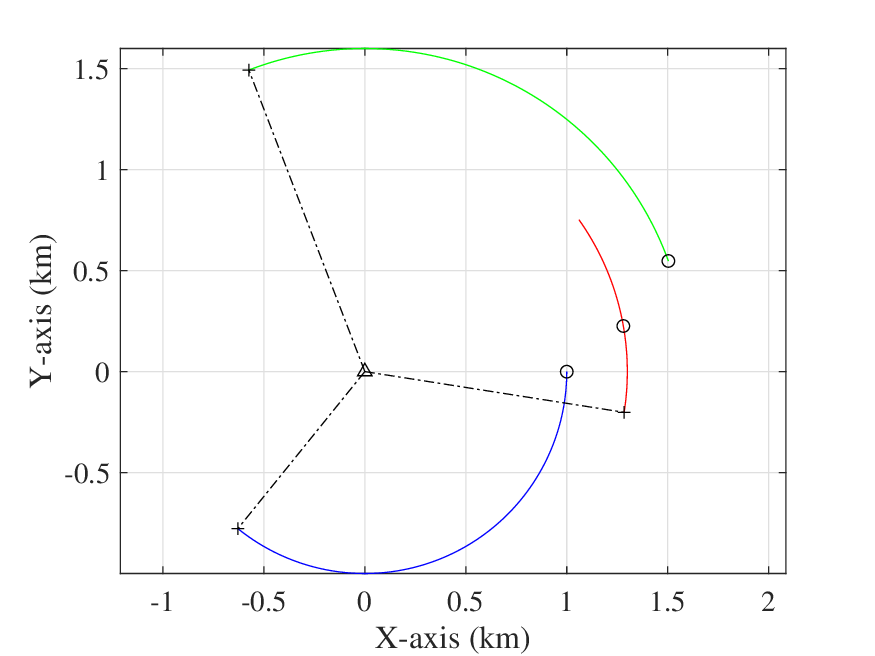}}
  \centerline{(a)}
\end{minipage}
\hfill
\begin{minipage}[b]{0.45\linewidth}
  \centering
  \centerline{\includegraphics[width=4.5cm]{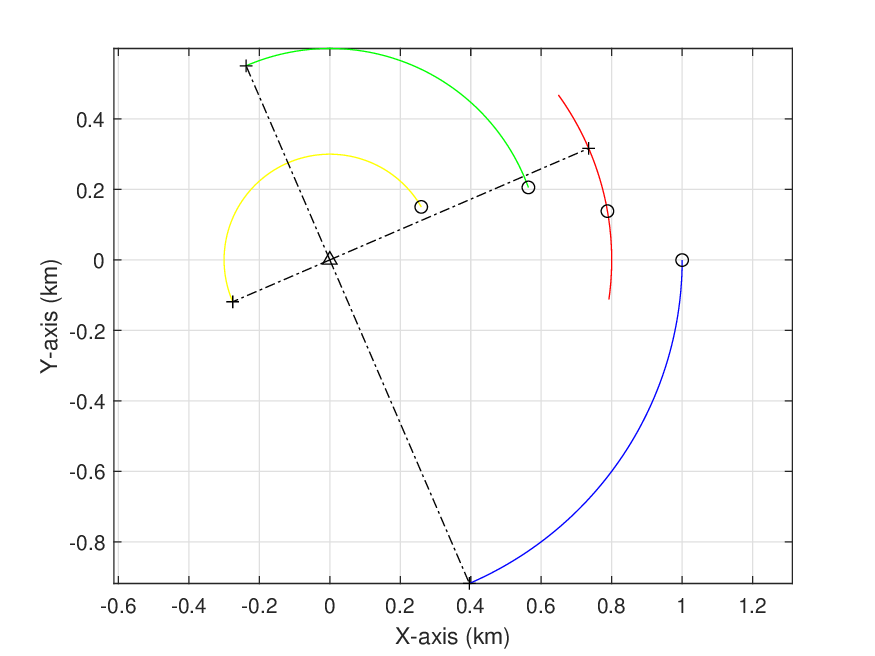}}
  \centerline{(b)}
\end{minipage}
\hfill
\begin{minipage}[b]{0.45\linewidth}
  \centering
  \centerline{\includegraphics[width=4.5cm]{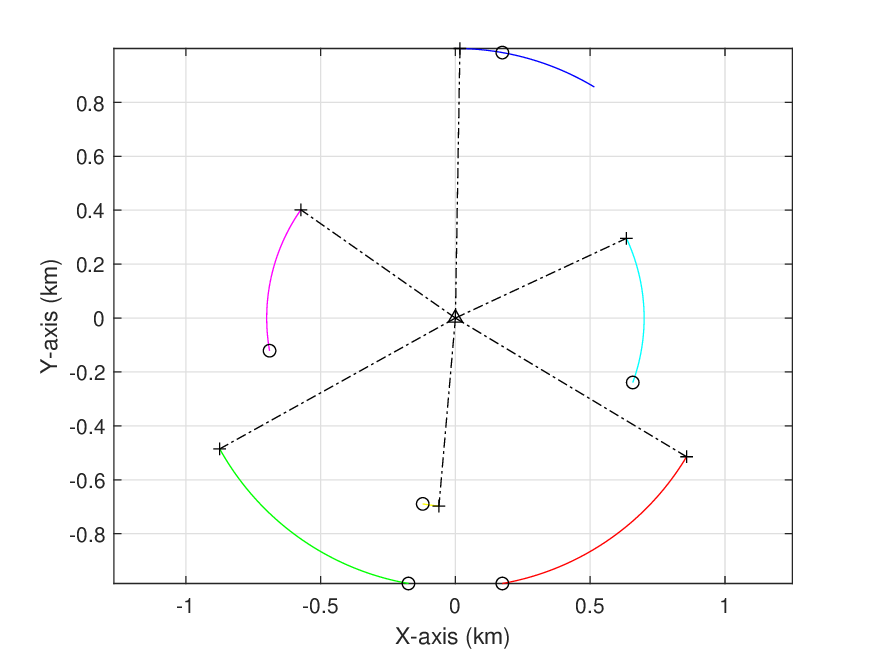}}
  \centerline{(c)}
\end{minipage}
\hfill
\begin{minipage}[b]{0.45\linewidth}
  \centering
  \centerline{\includegraphics[width=4.5cm]{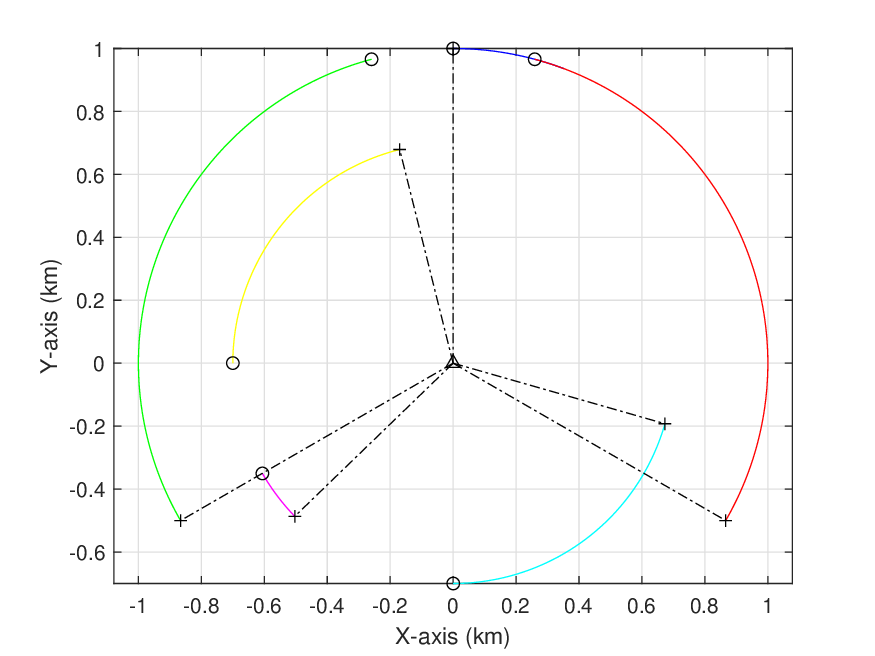}}
  \centerline{(d)}
\end{minipage}
\caption{Optimal sensor trajectories of two examples in Case 3 with nonuniform $d_{i}$. (a) nonuniform for $d_{1} = 1000$m, $d_{2} = 1300$m, and $d_{3} = 1600$m with 3 sensors. (b) nonuniform for $d_{1} = 300$m, $d_{2} = 600$m, $d_{3} = 800$m, and $d_{4} = 1000$m with 4 sensors. (b) two sensor groups with nonuniform $d_{1} = d_{2} = d_{3} = 1000$m and $d_{4} = d_{5} = d_{6} = 700$m. (d) two sensor groups with nonuniform $d_{1} = d_{2} = d_{3} = 1000$m and $d_{4} = d_{5} = d_{6} = 700$m.}
\label{fig_nonuni}
\end{figure}

\emph{Case 3 (nonuniform)}: Assume that $P_{0} = 1000$W, $\xi = 1$W/m, $\sigma = 0.5$m, $\rho = 1^{\circ}$,$\delta = 1$W, and $\gamma = 1.5$m. As shown in \textbf{Fig}. 4, in order to inspect the robustness of the strategies, the simulations are divided into four examples with different initial positions of the sensors. For \textbf{Fig}. 4(a), three sensors are utilized with random initial positions, i.e., $d_{1} = 1000$m, $d_{2} = 1300$m, $d_{3} = 1600$m. The sensors move from {$\{\alpha_{1} = 0^{\circ}, \;\alpha_{2} = 10^{\circ}, \;\alpha_{3} = 20^{\circ}\}$}. Though the sensor-source distances are in different values, the optimal geometries satisfy the UAA as well, and the optimal position is {$\{\alpha_{1} = 51.05^{\circ}, \;\alpha_{2} = 171.06^{\circ}, \;\alpha_{3} = -68.96^{\circ}\}$}. Similar results are demonstrated in \textbf{Fig}. 4(b) with four sensors. 

Next, two groups of sensors with different sensor-source distances are employed to search the optimal placement strategies shown in \textbf{Fig}. 4(c) and \textbf{Fig}. 4(d). Each group, including three sensors, has the same sensor-source distance, i.e., $d_{1} = d_{2} = d_{3} = 1000$m, $d_{4} = d_{5} = d_{6} = 700$m, and the initial position of sensors are {$\{\alpha_{1} = 80^{\circ}, \;\alpha_{2} = -80^{\circ}, \;\alpha_{3} = -100^{\circ}, \;\alpha_{4} = -20^{\circ}, \;\alpha_{5} = 190^{\circ}, \;\alpha_{6} = -100^{\circ}\}$} and {$\{\alpha_{1} = 90^{\circ}, \;\alpha_{2} = 75^{\circ}, \;\alpha_{3} = 105^{\circ}, \;\alpha_{4} = -90^{\circ}, \;\alpha_{5} = -150^{\circ}, \;\alpha_{6} = -180^{\circ}\}$}. \textbf{Fig}. 4(c) and \textbf{Fig}. 4(d) show that the UAA condition offers the optimal placement strategies for each group. The final geometries are {$\{\alpha_{1} = -90.97^{\circ}, \;\alpha_{2} = 149.03^{\circ}, \;\alpha_{3} = 29.04^{\circ}, \;\alpha_{4} = -155.01^{\circ}, \;\alpha_{5} = -35.01^{\circ}, \;\alpha_{6} = 84.99^{\circ}\}$} and {$\{\alpha_{1} = -89.99^{\circ}, \;\alpha_{2} = 150.01^{\circ}, \;\alpha_{3} = 30.00^{\circ}, \;\alpha_{4} = 164.05^{\circ}, \;\alpha_{5} = 44.05^{\circ}, \;\alpha_{6} = -75.95^{\circ}\}$}

\section{Conclusion}
\label{sec:conc}

In this paper, we have developed the optimal sensor placement strategies using combinations of hybrid measurements. Three contributions have been made. Firstly, the minimum tr(CRB)s of different combinations of TDOA, AOA, RSS, and TOA are derived. Secondly, the optimal geometries are proposed to achieve the minimum tr(CRB)s. We found that the analytical solutions can be summarised into two groups. Thirdly, the constraints corresponding to specific measurements are revealed and the unified constraints of combinations are given. Simulation examples have verified the findings in this paper. Our future work will focus on developing algebraic algorithms to quickly calculate the optimal sensor-source geometries for general cases with diverse sensor measurement noise variances.



\begin{thebibliography}{1}
\bibliographystyle{IEEEtran}

\bibitem{c23} P. Khomchuk, R. S. Blum and I. Bilik, “Performance analysis of target parameters estimation using multiple widely separated antenna arrays," \textit{IEEE Trans. Aero. and Elec. Sys.}, vol. 52, no. 5, pp. 2413-2435, 2016.

\bibitem{c24} T. Tirer and A. J. Weiss, “High resolution direct position determination of radio frequency sources," \textit{IEEE Signal Processing Letters}, vol. 23, no. 2, pp. 192-196, 2016.

\bibitem{Kong2021} J. Wakulicz, H. Kong, and S. Sukkarieh, “Active information acquisition under arbitrary unknown disturbances," \textit{Proc. of 2021 IEEE Int. Conf. on Robotics and Automation (ICRA)}, pp. 8429-8435, 2021.

\bibitem{Zhou2023IROs} P. Li and L. Zhou, “Assignment algorithms for multi-robot multi-target tracking with sufficient and limited sensing capability," \textit{Proc. of IEEE/RSJ Int. Conf. on Intelligent Robots and Systems (IROS)}, pp. 1-8, 2023.

\bibitem{Zhou2023ACC} R. Zahroof, J. Liu, L. Zhou, and V. Kumar , “Multi-robot localization and target tracking with connectivity maintenance and collision avoidance," \textit{Proc. of American Control Conference (ACC)}, pp. 1331-1338, 2023.

\bibitem{Zhou2023TRO} L. Zhou and V. Kumar, “Robust multi-robot active target tracking against sensing and communication attacks," \textit{IEEE Trans. Robotics}, vol. 39, no. 3, pp. 1768-1780, 2023.

\bibitem{c1} B. Huang, L. Xie, and Z. Yang, “TDOA-based source localization with distance-dependent noises,” \textit{IEEE Trans. Wireless Commun.}, vol. 14, no. 1, pp. 468–480, 2015.

\bibitem{Su2021} D. Su, H. Kong, S. Sukkarieh and S. Huang, “Necessary and sufficient conditions for observability of SLAM-based TDOA sensor array calibration and source localization", \textit{IEEE Trans. Robotics}, Vol. 37, No. 5, pp. 1451-1468, 2021. 

\bibitem{He2023} Y. He, J. Wang, D. Su, K. Nakadai, J. Wu, S. Huang, Y. Li, and H. Kong, “Observability analysis of graph SLAM-based joint calibration of multiple microphone arrays and sound source localization“, \textit{Proc. of the IEEE/SICE SII}, pp. 1-8, 2023. 

\bibitem{c2} J. Shen, A. F. Molisch, and J. Salmi, “Accurate passive location estimation using TOA measurements,” \textit{IEEE Trans. Wireless Commun.}, vol. 11, no. 6, pp. 2182–2192, 2012.

\bibitem{c3} A. J. Weiss, “On the accuracy of a cellular location system based on RSS measurements,” \textit{IEEE Trans. Veh. Technol.}, vol. 52, no. 6, pp. 1508–1518, 2003.

\bibitem{Kong2023} L. Fu, X. Qiao, S. Huang, G. Mao, Z. Lin, Y. Li, and H. Kong, “SLAM-based joint calibration of differential RSS
sensor array and source localization,” \textit{Proc. of the 49th Annual Conference of the IEEE Industrial Electronics Society (IECON)}, pp. 1–8, 2003.

\bibitem{c4} Y. Zhu, D. Huang, and A. Jiang, “Network localization using angle of arrival,” \textit{Proc. IEEE Int. Conf. Electro/Inf. Technol.}, pp. 205–210, 2008.

\bibitem{c6} K. Panwar, M. Katwe, P. Babu, P. Ghare, and K. Singh, “A majorization-minimization algorithm for hybrid TOA-RSS based localization in NLOS environment,” \textit{IEEE Commun. Lett.}, vol. 26, no. 5, pp. 1017–1021, 2022.

\bibitem{c7} S. Chang, Y. Li, X. Yang, H. Wang, W. Hu, and Y. Wu, “A novel localization method based on RSS-AOA combined measurements by using polarized identity,” \textit{IEEE Sensors J.}, vol. 19, no. 4, pp. 1463–1470, 2019.

\bibitem{c8} V. Y. Zhang, A. K.-s.Wong, K. T.Woo, and R.W. Ouyang, “Hybrid TOA/AOA-based mobile localization with and without tracking in CDMA cellular networks,” \textit{Proc. IEEE Wireless Commun. Netw. Conf.}, pp. 1–6, 2010.

\bibitem{c5} Y. Wang and K. C. Ho, “Unified near-field and far-field localization for AOA and hybrid AOA-TDOA positionings,” \textit{IEEE Trans. Wireless Commun.}, vol. 17, no. 2, pp. 1242–1254, 2018.

\bibitem{c9} J. Fawcett, “Effect of course maneuvers on bearings-only range estimation,” \textit{IEEE Trans. Acoust., Speech, Signal Process.}, vol. 36, no. 8, pp. 1193–1199, 1988.

\bibitem{c10} B. Yang and J. Scheuing, “Cramer-Rao bound and optimum sensor array for source localization from time differences of arrival,” \textit{Proc. Int. Conf. Acoust., Speech, Signal Process.}, 2005, vol. 4, 2005.

\bibitem{c11} K. Do{\u{g}}an{\c{c}}ay and H. Hmam, “Optimal angular sensor separation for AOA localization,” \textit{Signal Process.}, vol. 88, no. 5, pp. 1248–1260, 2008.

\bibitem{c12} S. Xu and K. Do{\u{g}}an{\c{c}}ay, “Optimal sensor placement for 3-D angle-of- arrival target localization,” \textit{IEEE Trans. Aerosp. Electron. Syst.}, vol. 53, no. 3, pp. 1196–1211, 2017.

\bibitem{c13} N. H. Nguyen and K. Do{\u{g}}an{\c{c}}ay, “Optimal geometry analysis for multistatic TOA localization,” \textit{IEEE Trans. Signal Process.}, vol. 64, no. 16, pp. 4180–4193, 2016.

\bibitem{c14} S. Xu, Y. Ou, and W. Zheng, “Optimal sensor-target geometries for 3-D static target localization using received-signal-strength measurements,” \textit{IEEE Signal Process. Lett.}, vol. 26, no. 7, pp. 966–970, 2019.

\bibitem{c15} S. Xu, Y. Ou, and X. Wu, “Optimal sensor placement for 3-D time-of-arrival target localization,” \textit{IEEE Trans. Signal Process.}, vol. 67, no. 19, pp. 5018–5031, 2019.

\bibitem{c16} N. Sahu, L.Wu, P. Babu, B. Shankar, and B. Ottersten, “Optimal sensor placement for source localization: A unified ADMM approach,” \textit{IEEE Trans. Veh. Technol.}, vol. 71, no. 4, pp. 4359–4372, 2022.

\bibitem{c17} C. Yang, L. Kaplan, E. Blasch, and M. Bakich, “Optimal Placement of Heterogeneous Sensors for Targets with Gaussian Priors," \textit{IEEE Trans. Aerosp. Electron. Syst.}, vol. 49(3), pp. 1637-1653, 2013.

\bibitem{c18} W. Meng, L. Xie, and W. Xiao, “Optimality analysis of sensor-source geometries in heterogeneous sensor networks," \textit{IEEE Trans. Wireless Commun.}, vol. 12(4), pp. 1958-1967, 2013.

\bibitem{c19} S. Xu, “Optimal sensor placement for target localization using hybrid RSS, AOA and TOA measurements,” \textit{IEEE Commun. Lett.}, vol. 24, no. 9, pp. 1966–1970, 2020.

\bibitem{c20} A. N. Bishop, B. Fidan, B. D.O. Anderson, and K. Doğançay, P. N. Pathirana, “Optimality analysis of sensor-target localization geometries," \textit{Automatica}, vol. 46, no. 3, pp. 479-492, 2010.

\bibitem{c21} D. Ucinski, \textit{Optimal measurement methods for distributed parameter system identification}. Boca Raton, FL, USA: CRC Press, 2005.

\bibitem{c22} W. Wang, P. Bai, X. Liang, Y. Wang, and J. Zhang, “Optimal deployment of sensor–emitter geometries for hybrid localisation using TDOA and AOA measurements," \textit{IET Sci. Meas. Technol.}, vol. 13, pp. 622-631, 2019.

\end{thebibliography}
\end{document}